\documentstyle[emulateapj,apjfonts,epsfig,graphicx,natbib,bib]{article}
\def\etal{{et al.}}
\def\meszaros{M\'{e}sz\'{a}ros}

\begin{document}

   \title{Jet Break Time -- Flux Density Relationship and Constraints on Physical Parameters of Gamma-Ray Burst Afterglows}

\author{X. F. Wu$^{1}$, Z. G. Dai$^{1}$, and E. W. Liang$^{1,2,3}$}
\affil{$^1$Department of Astronomy, Nanjing University, Nanjing
210093, China; Email:xfwu@nju.edu.cn, dzg@nju.edu.cn, ewliang@nju.edu.cn.\\
$^2$Department of Physics, Guangxi University, Nanning 530004, China.\\
$^3$National Astronomical Observatories, Yunnan Observatory,
Chinese Academy of Sciences, Kunming 650011, China.}

\begin{abstract}
We derive a relation between the flux density $F_{\nu,j}$ at the
light-curve break of a gamma-ray burst (GRB) afterglow and the
break time $t_{j}$. The break is due to the transition from the
spherical-like to jet-like evolution of the afterglow, when the
Lorentz factor of the jet equals the inverse of the initial
half-opening angle, i.e., $\gamma=1/\theta_0$. We show that this
relation indeed behaves as $F_{\nu,j}\propto t_{j}^{-p}$ among
GRBs for the slow-cooling case, where $p$ is the power-law index
of electron distribution. A statistical analysis of the optical
jet breaks of nine GRBs gives $p=2.10\pm 0.21$, which is
consistent with the shock acceleration theory. The value of $p$
derived in this way is different from the observed temporal index
$\alpha_2$ ($F_{\nu}\propto t^{-\alpha_{2}}$) of the late-time
light curve after $t_{j}$, which suffers several uncertainties
from the unclear hydrodynamics of the sideways expansion and
exhibits a large dispersion. Our results not only confirm that the
remnants of GRBs are standard candles, but also provide the first
evidence that the physical parameters of relativistic shocks are
universal, with the favored values $\epsilon_{e}\sim 0.1$ and
$\epsilon_{B}\sim 10^{-3}$.

\end{abstract}

\section{Introduction}
\label{sec:introduction} The gamma-ray burst (GRB) afterglows are
attributed to the nonthermal synchrotron/inverse Compton (IC)
radiation from the swept-up circumburst electrons shocked by
relativistic blast waves (\citealt{wrm+97,wax97,kp+97}). There are
two popular types of circumburst medium, i.e., the interstellar
medium (ISM) and the stellar wind (for the latter see
\citealt{dl+98,mrw+98,cl+99}). Nevertheless, the ambient electrons
are initially accelerated to a power-law distribution in the same
way, $dN/d\gamma_{e}\propto\gamma_{e}^{-\it{p}}$
($\gamma_{m}\leq\gamma_{e}\leq\gamma_{\max}$), with the typical
index $p\sim 2.2-2.3$ (Achterberg et al. 2001 and references
therein; Lemoine \& Pelletier 2003). The minimum Lorentz factor
$\gamma_{m}$ is proportional to the bulk Lorentz factor $\gamma$
of the shock and the energy equipartition factor $\epsilon_{e}$ of
the electrons. Magnetic fields can also be generated by the shock
through the relativistic Weibel instability, with the energy
equipartition factor $\epsilon_{B}$ of $10^{-5}$ to $10^{-1}$
(\citealt{ml+99}). The postshock electrons with Lorentz factor
$\gamma_{c}$ will convenienly lose their total energy in the
dynamical timescale because of synchrotron and IC radiation
(\citealt{spn+98}). The initial distribution of the electrons is
thus approximated by a broken power law.

It is now the consensus of most GRB researchers that the GRB
fireball is not spherical but indeed conical or jetted.
\cite{fks+01} established the ``standard candle'' hypothesis of
geometrically corrected gamma-ray energy release ($E_{\gamma}\sim
5\times 10^{50}$ ergs) of prompt GRBs based on the previous work
of \cite{rhoads99} and \cite{sph+99} on the hydrodynamic evolution
of a relativistic jet (see also \citealt{bfk+03}).
\cite{pk+01,pk+02} have performed multiwavelength fitting to $10$
GRB afterglows and given a comparable mean energy in the jets at
the afterglow stage. Statistics of the late-time X-ray luminosity
of GRBs further confirms the standard energy outputs in GRB
afterglows (\citealt{fw+01,pkpp+01,bkf+03}). It also requires the
small scatter of $p$ with the mean value $p\approx 2$
(\citealt{bkf+03}).

In this paper we investigate the energetics of GRB afterglows and
the physical parameters related to relativistic shock physics by
studying the light curve breaks of GRB optical transients (OTs) in
the statistical sense. We derive the analytical relation between
the flux density and time at the break in \S2. We list the sample
and give our statistical results in \S3. The findings and
implications of our work are discussed in \S4.

\section{Spectral properties at the jet break time}
\label{sec:relation} The observed synchrotron spectrum can be
determined by the typical frequency $\nu_{m}$ corresponding to the
electrons with Lorentz factor $\gamma_{m}$, the cooling frequency
$\nu_{c}$ corresponding to the electrons with Lorentz factor
$\gamma_{c}$, and the peak flux density $F_{\nu,\max}$. To
calculate these quantities, we assume an adiabatic jet with
initial half-opening angle $\theta_{0}$. At earlier times, the
bulk Lorentz factor $\gamma$ of the jet is larger than
$\theta_{0}^{-1}$, and its radiation shows no difference from that
of an isotropic fireball. The light curve steepens achromatically
when $\gamma\leq\theta_{0}^{-1}$, because of the deficit of the
radiating area for a nonlateral expansion jet or the ultimate
change of the hydrodynamics for a lateral-expansion jet. Here we
focus on the transitional moment $t_{j}$ when
$\gamma=\theta_{0}^{-1}$ (\citealt{sph+99}). The emission
properties of an isotropic fireball can be applied to this time,
and we derive the flux density $F_{\nu,j}$ $[=F_{\nu}(t_{j})]$ as
a function of $t_{j}$ for both the ISM case and the stellar wind
case.

\subsection{The ISM case}
\label{sec:relation-ism} For the ISM case (e.g., Sari et al.
1998), we have
\begin{equation}
t_{j}=0.82(1+z)E_{j,51}^{1/3}n^{-1/3}\theta_{0,-1}^2\,\,{\rm{days,}}
\end{equation}
\begin{equation}
\nu_{m,j}=2.7\times10^{11}\kappa_{m}(1+z)^{-1}\epsilon_{e,-1}^2\epsilon_{B,-3}^{1/2}\zeta_{1/6}^2n^{1/2}\theta_{0,-1}^{-4}\,\,{\rm{Hz,}}
\end{equation}
\begin{equation}
\nu_{c,j}=2.1\times10^{16}\kappa_{c}(1+z)^{-1}\epsilon_{B,-3}^{-3/2}E_{j,51}^{-2/3}n^{-5/6}(1+Y_{j})^{-2}\,\,{\rm{Hz,}}
\label{eqn:nuc-ism}
\end{equation}
\begin{equation}
F_{\nu,\max,j}=70\kappa_{f}(1+z)\epsilon_{B,-3}^{1/2}E_{j,51}\theta_{0,-1}^{-2}n^{1/2}D_{L,28}^{-2}\,\,{\rm{mJy,}}
\end{equation}
where $z$ is the redshift, $D_{L}$ is the luminosity distance,
$\zeta_{1/6}=6(p-2)/(p-1)$,
$E_{j}\approx\frac{1}{2}E_{{\rm{iso}}}\theta_{0}^2$ is the total
jet energy, and $n$ is the density of the ISM in units of
cm$^{-3}$. We adopt the convention $Q=10^{x} Q_{x}$. We have
considered here the accurate expressions for $\nu_{m}$, $\nu_{c}$,
and $F_{\nu,\max}$, based on the self-similar solutions of
\cite{bm+76} for the spherical blast waves performed by
\cite{gs+02}. The Granot \& Sari (2002) corrections to the
formulae of \cite{spn+98} are denoted as
$\kappa_{m}=0.73(p-0.67)$,
$\kappa_{c}=(p-0.46)\exp{(3.16-1.16p)}$, and
$\kappa_{f}=0.09(p+0.14)$. These factors are nearly constant for
$p$ in the range of $2.0-2.5$, and $\kappa_{m}=1.12$,
$\kappa_{c}=3.22$, and $\kappa_{f}=0.21$ for $p=2.2$. The Compton
parameter $Y_{j}=Y(t_{j})$ is mainly determined by the ratio
$\epsilon_{e}/\epsilon_{B}$ (\citealt{pk+00,se+01}). It can be
neglected in equation (\ref{eqn:nuc-ism}) if
$\epsilon_{e}/\epsilon_{B}\lesssim 1$. However, when
$\epsilon_{e}/\epsilon_{B}\gg 1$, $\nu_{c,j}$ is rewritten as
\begin{eqnarray}
\nu_{c,j}&=&4.5\times10^{14}\kappa_{c}(1+z)^{-1}[1716^{p-2.2}(\frac{\kappa_{c}}{\kappa_{m}})^{p-2}\epsilon_{e,-1}^{2(1-p)}\epsilon_{B,-3}^{-p/2}\zeta_{1/6}^{2(2-p)}\nonumber\\
              & &\times E_{j,51}^{-4/3}\theta_{0,-1}^{4(p-2)}n^{-(3p+4)/6}]^{1/(4-p)}\,\,{\rm{Hz,}}
\end{eqnarray}
where the electrons are assumed to be in the IC-dominated slow
cooling case at $t_{j}$ (\citealt{se+01}). The transition time
from fast cooling to IC-dominated slow-cooling
$t_0^{\rm{IC}}=(\epsilon_{e}/\epsilon_{B})t_0=5\times
10^{-4}(\kappa_{m}/\kappa_{c})(1+z)\epsilon_{e,-1}^3
\epsilon_{B,-3}\zeta_{1/6}^{2}E_{\rm{iso},53}n$ days, measured by
the observer, is earlier than typical break time $t_{j}\sim 1$ day
(see Table ~\ref{tab:break}), while the moment $t^{\rm{IC}}$ when
the synchrotron cooling begins to dominate over the IC cooling is
typically more than years after the GRB trigger (\citealt{se+01}).

The flux density at the jet break time in the slow-cooling
spectrum case ($\nu_{m}<\nu<\nu_{c}$) is
\begin{eqnarray}
 F_{\nu,j}&=&70t_{j,\rm{day}}^{-p}\times73.7^{2.2-p}\kappa_{f}\kappa_{m}^{(p-1)/2}\epsilon_{e,-1}^{p-1}\epsilon_{B,-3}^{(p+1)/4}\zeta_{1/6}^{p-1}n^{(3-p)/12}\nonumber\\
               & &\times E_{j,50.5}^{(p+3)/3}D_{L,28}^{-2}(1+z)^{(p+3)/2}(\frac{\nu}{\nu_R})^{-(p-1)/2}\,\,{\rm{\mu Jy}},
\label{eqn:sc-break-ism}
\end{eqnarray}
where $\nu_{R}=4.55\times10^{14}$ Hz is the $R$-band frequency
taken as the observed frequency. Equation (\ref{eqn:sc-break-ism})
provides a relationship between the flux density $F_{\nu,j}$ and
the jet break time $t_{j}$. In the following this relationship is
called the jet break relation. On the other hand, a similar
relation in the fast-cooling case ($\nu_{c}<\nu$) follows
\begin{eqnarray}
 F_{\nu,j}&=&700t_{j,\rm{day}}^{-p}\times73.7^{2.2-p}\kappa_{f}\kappa_{m}^{(p-1)/2}\kappa_{c}^{1/2}D_{L,28}^{-2}\epsilon_{e,-1}^{p-1}\epsilon_{B,-3}^{(p-2)/4}\zeta_{1/6}^{p-1}\nonumber\\
               & &\times E_{j,50.5}^{(p+2)/3}n^{-(p+2)/12}(1+z)^{(p+2)/2}(1+Y_{j})^{-1}(\frac{\nu}{\nu_R})^{-p/2}\,\,{\rm{\mu Jy}},
\end{eqnarray}
which can be rewritten as
\begin{eqnarray}
 F_{\nu,j}&=&114t_{j,\rm{day}}^{-p+(p-2)/(4-p)}\times73.7^{2.2-p}\kappa_{f}\kappa_{m}^{p/2}D_{L,28}^{-2}(\frac{\nu}{\nu_R})^{-p/2} \nonumber\\
               & &\times[121^{p-2.2}\frac{\kappa_{c}}{\kappa_{m}}\epsilon_{e,-1}^{2p-p^2-3}\epsilon_{B,-3}^{(4+2p-p^2)/4}\zeta_{1/6}^{4p-p^2-2}E_{j,50.5}^{(12-p^2)/3}\nonumber\\
               & &\times n^{p(p-6)/12}(1+z)^{(12-p^2)/2}]^{1/(4-p)}\,\,{\rm{\mu Jy}},
\label{eqn:fc-break-ism}
\end{eqnarray}
in the limit of $\epsilon_{e}/\epsilon_{B}\gg 1$, where we do not
include the contribution of the synchrotron-self-Compton (SSC)
scattering to the flux density, since the SSC component always
appears above the X-ray band for typical physical parameters and
the light curve breaks are mostly observed in optical band.

\subsection{The stellar wind case}
\label{sec:relation-wind} For the stellar wind case (e.g.,
Chevalier \& Li 2000), we have
\begin{equation}
t_{j}=1.25(1+z)E_{j,51}A_{\ast}^{-1}\theta_{0,-1}^2\,\,{\rm{days,}}
\end{equation}
\begin{equation}
\nu_{m,j}=2.9\times10^{11}\kappa_{m}(1+z)^{-1}\epsilon_{e,-1}^2\epsilon_{B,-3}^{1/2}\zeta_{1/6}^2E_{j,51}^{-1}A_{\ast}^{3/2}\theta_{0,-1}^{-4}\,{\rm{Hz,}}
\end{equation}
\begin{equation}
\nu_{c,j}=2.8\times10^{16}\kappa_{c}(1+z)^{-1}\epsilon_{B,-3}^{-3/2}E_{j,51}A_{\ast}^{-5/2}(1+Y_{j})^{-2}\,\,{\rm{Hz,}}
\end{equation}
\begin{equation}
F_{\nu,\max,j}=7.45\kappa_{f}(1+z)\epsilon_{B,-3}^{1/2}A_{\ast}^{3/2}\theta_{0,-1}^{-2}D_{L,28}^{-2}\,\,{\rm{mJy,}}
\end{equation}
where $A_{\ast}$ is the wind parameter (\citealt{cl+99}). The
correction factors derived from \cite{gs+02} are
$\kappa_{m}=0.4(p-0.69)$, $\kappa_{c}=(3.45-p)\exp{(0.45p-1.4)}$,
and $\kappa_{f}=1.31(p+0.12)$. For $p=2.2$, $\kappa_{m}=0.60$,
$\kappa_{c}=0.83$, and $\kappa_{f}=3.04$. When
$\epsilon_{e}/\epsilon_{B}\gg 1$, the transition time from fast
cooling to IC-dominated slow cooling in the wind case is
$t_0^{\rm{IC}}=(\epsilon_{e}/\epsilon_{B})^{1/2}t_0=4\times
10^{-2}(\kappa_{m}/\kappa_{c})^{1/2}(1+z)\epsilon_{e,-1}^{3/2}
\epsilon_{B,-3}^{1/2}\zeta_{1/6}A_{\ast}$ days, which is earlier
than $t_{j}\sim 1$ day. In this case $\nu_{c,j}$ is rewritten as
\begin{eqnarray}
\nu_{c,j}&=&6.0\times10^{14}\kappa_{c}(1+z)^{-1}[2151^{p-2.2}(\frac{\kappa_{c}}{\kappa_{m}})^{2(p-2)}\epsilon_{e,-1}^{2(1-p)}\epsilon_{B,-3}^{-p/2}\zeta_{1/6}^{2(2-p)}\nonumber\\
              & &\times E_{j,51}^{p}\theta_{0,-1}^{4(p-2)}A_{\ast}^{-(3p+4)/2}]^{1/(4-p)}\,\,{\rm{Hz.}}
\end{eqnarray}

The flux density at the jet break time in the slow-cooling case
($\nu_{m}<\nu<\nu_{c}$) is
\begin{eqnarray}
 F_{\nu,j}&=&23t_{j,\rm{day}}^{-p}\times57.3^{2.2-p}\kappa_{f}\kappa_{m}^{(p-1)/2}\epsilon_{e,-1}^{p-1}\epsilon_{B,-3}^{(p+1)/4}\zeta_{1/6}^{p-1}A_{\ast}^{(3-p)/4}\nonumber\\
               & &\times E_{j,50.5}^{(p+1)/2}D_{L,28}^{-2}(1+z)^{(p+3)/2}(\frac{\nu}{\nu_R})^{-(p-1)/2}\,\,{\rm{\mu Jy}},
\label{eqn:sc-break-wind}
\end{eqnarray}
while the density flux at $\nu_{c}<\nu$ becomes
\begin{eqnarray}
 F_{\nu,j}&=&100t_{j,\rm{day}}^{-p}\times57.3^{2.2-p}\kappa_{f}\kappa_{m}^{(p-1)/2}\kappa_{c}^{1/2}D_{L,28}^{-2}\epsilon_{e,-1}^{p-1}\epsilon_{B,-3}^{(p-2)/4}\zeta_{1/6}^{p-1}\nonumber\\
               & &\times E_{j,50.5}^{(p+2)/2}A_{\ast}^{-(p+2)/4}(1+z)^{(p+2)/2}(1+Y_{j})^{-1}(\frac{\nu}{\nu_R})^{-p/2}\,{\rm{\mu Jy}},
\end{eqnarray}
which can be further deduced in the case of
$\epsilon_{e}/\epsilon_{B}\gg 1$,
\begin{eqnarray}
 F_{\nu,j}&=&14.5t_{j,\rm{day}}^{-p+(p-2)/(4-p)}\times57.3^{2.2-p}\kappa_{f}\kappa_{m}^{p/2}D_{L,28}^{-2}(\frac{\nu}{\nu_R})^{-p/2} \nonumber\\
               & &\times E_{j,50.5}^{(p+2)/2}[36.2^{p-2.2}(\frac{\kappa_{c}}{\kappa_{m}})^{p/2}\epsilon_{e,-1}^{2p-p^2-3}\epsilon_{B,-3}^{(4+2p-p^2)/4}\zeta_{1/6}^{4p-p^2-6}\nonumber\\
               & &\times A_{\ast}^{p(p-6)/4}(1+z)^{(12-p^2)/2}]^{1/(4-p)}\,\,{\rm{\mu Jy}}.
\label{eqn:fc-break-wind}
\end{eqnarray}

We can see that in the slow-cooling case, the jet break relation
behaves as $F_{\nu,j}\propto t_{j}^{-p}$ among GRB afterglows, as
long as the physical parameters $E_{j}$, $\epsilon_{e}$, and
$\epsilon_{B}$ are clustered. This relation is assured in both the
ISM and the stellar wind case, and is insensitive to the medium
density in each case. It provides a tool for probing the
energetics of GRB afterglows and the shock physics of relativistic
blast waves. The jet break relation of the fast-cooling case is
affected by the Compton parameter $Y_{j}$. If $Y_{j}<1$, the jet
break relation becomes $F_{\nu,j}\propto t_{j}^{-p}$ and is
insensitive to $\epsilon_{B}$ while moderately sensitive to the
ambient medium density. If $Y_{j}>1$, the relation changes to
$F_{\nu,j}\propto t_{j}^{-p+(p-2)/(4-p)}$, in which the index
$p-(p-2)/(p-4)$ is in the range of $2.0-2.17$ for $p\sim 2.0-3.0$.

\section{Statistical results}
\label{sec:statistics}
\subsection{The sample}
\label{sec:sample} In Table~\ref{tab:break} we give an updated
optical $R$-band sample of $14$ GRB light curve breaks. Our sample
is slightly different from the Bloom et al. (2003; hereafter
BFK03) sample. In the BFK03 sample of 17 GRBs with known $t_{j}$,
they included six jet breaks that were determined from
observations outside the optical bands, e.g., GRBs $970508$,
$980703$, and $000418$ at the radio band; GRB $990705$ at $H$
band; GRB $010921$ at the joint $I$ and F$702$ W bands; and GRB
$970828$ at the X-ray band. Recently, two new $R$-band jet breaks
of GRBs $030226$ and $030329$ have been observed. Greiner et al.
(2003a) concluded that a jet break existed in the $R$-band light
curve of the GRB $011121$ afterglow before $10$ days. We also add
this GRB in Table~\ref{tab:break}. Since the error of $t_{j}$ of
GRB $011121$ is large, it will not significantly change our
statistical results. There are two peculiar OTs in our sample. GRB
$000301$C showed a prebreak bump in the optical afterglow that has
been explained as being caused by the central engine activity, by
the external density jump, or by the microlensing event
(\citealt{bc+00,dl+01,dl+02,gls+00}). GRB $021004$ also exhibited
complicated fluctuations in the early afterglow before the
temporal break (\citealt{fyk+03}).

In Table~\ref{tab:fastfading} we give a sample of five fast-fading
GRB optical afterglows. Fast-fading afterglows are believed to be
already in the jet-like stages before they are definitely
observed. The upper limit of $t_{j}$ is the time when the first
positive optical detection was made. We exclude GRB $980329$ in
the BFK03 sample because the temporal index is relatively too
shallow to be identified as a fast-fading GRB ($\alpha=1.28\pm
0.19$; \citealt{rlm+99}). The optical afterglow of GRB $990705$
may be a fast-fading one, but there is no reliable $R$-band data
for this GRB.

\subsection{Results}
\label{sec:results} From Table~\ref{tab:break} we can see that the
optical spectral index $\beta_{\rm{opt}}$ around $t_{j}$ is less
than $0.8$ for most GRB OTs. This can be interpreted as the
optical frequency located at the slow-cooling segment
($\nu_{m}<\nu<\nu_{c}$) of the spectrum, since the observed index
is consistent with the theoretical one, $\beta=(p-1)/2\sim
0.6-0.75$ for $p\sim 2.2-2.5$. There are two outliers, GRB
$000926$ and $020813$, which have relatively steep spectra at
$t_{j}$ and can be regarded to be in the fast-cooling case in
which the typical spectral index $\beta=p/2\sim1.1$. We thus adopt
the jet break relation of the slow-cooling case for the
statistical purpose in this work.

To decouple the effects of the redshift and the luminosity
distance from other parameters, we rewrite equations
(\ref{eqn:sc-break-ism}) and (\ref{eqn:sc-break-wind}), and the
general jet break relation in the slow-cooling case is
\begin{equation}
\mathcal{L_{J}}=\it{a-b\tau_{j}}, \label{eqn:sc-break}
\end{equation}
where $\mathcal{L_{J}}{\equiv
\log[\it{F_{\nu,j}D_{L,\rm{28}}^{\rm{2}}(\rm{1}+\it{z})^{\rm{-3/2}}}/\mu\rm{Jy}]}$
and $\tau_{j}\equiv \log[t_{j}(1+z)^{-1/2}/\rm{days}]$ are the
equivalent luminosity density and jet break time in logarithms,
which can be determined directly by observations. The coefficient
$a$ is a combination of the physical parameters of the central
engine and the shock physics and is insensitive to the external
medium density. However, $b\equiv p$ is only determined by the
index of the distribution of the electrons. We adopt the cosmology
with $\Omega_{m}=0.27$, $\Omega_{\Lambda}=0.73$, and $H_{0}=71$ km
s$^{-1}$ Mpc$^{-1}$.

\begin{figure}
\plotone{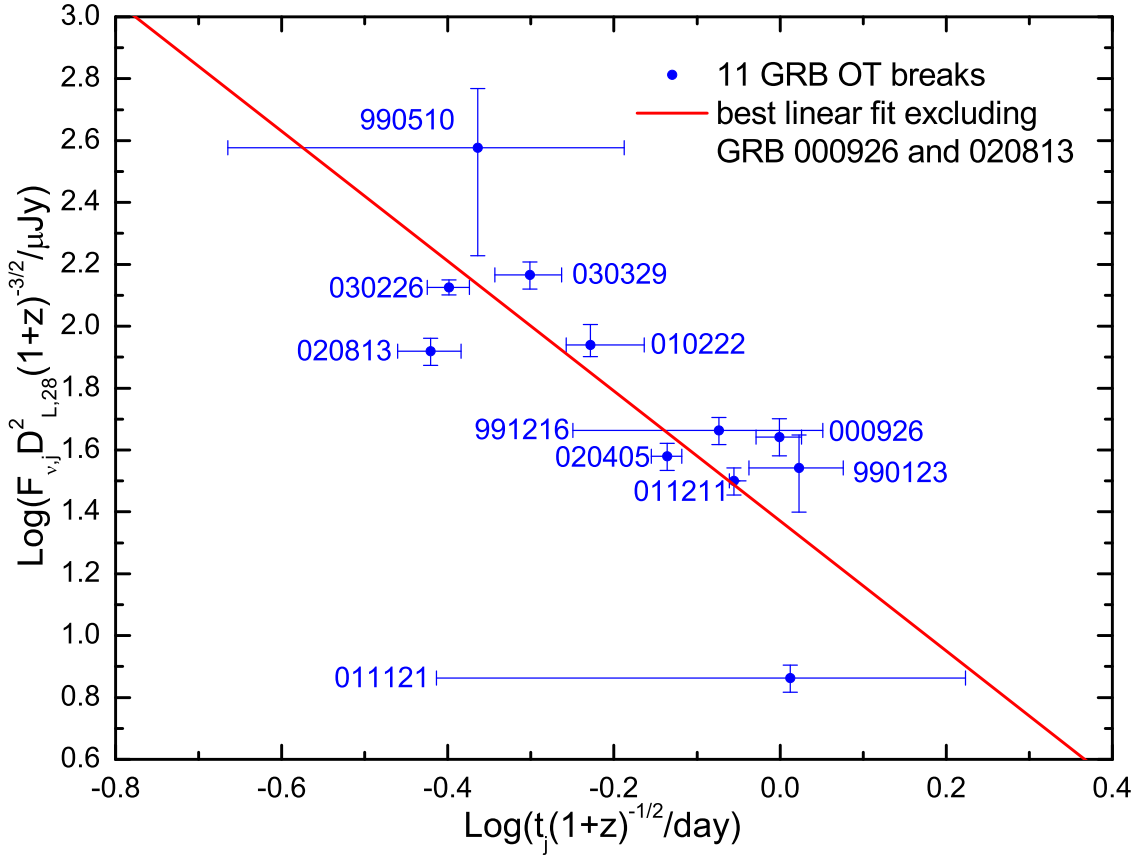} \caption{Plot of $\mathcal{L_{J}}$ as a
function of $\tau_{j}$ for $11$ well-observed GRB $R$-band jet
break data with known redshifts listed in Table~\ref{tab:break},
excluding two peculiar GRB OTs (GRB $000310$C and $021004$). The
solid line shows the best linear fit to these data for the
slow-cooling jet break relation (i.e., eqs.
[\ref{eqn:sc-break-ism}] and [\ref{eqn:sc-break-wind}]):
$a=1.37\pm 0.05$, $b=2.10\pm 0.21$ ($1 \sigma$), $\chi^2=9.97$ for
$9-2$ degrees of freedom, and the possibility $Q(>\chi^2)=0.19$.
The jet breaks of GRBs $000926$ and $020813$ are considered to be
the fast-cooling ones by their spectra and therefore are not
included in the linear fit. \label{fig:11-breaks}}
\end{figure}

Figure \ref{fig:11-breaks} shows the $\mathcal{L_{J}}$ -
$\tau_{j}$ plot for $11$ GRBs with known redshifts. Excluding GRBs
$000926$ and $020813$ as explained above, we have made the best
linear fit to the remaining nine GRB OT breaks using equation
(\ref{eqn:sc-break}).\footnotemark\footnotetext{\label{foot:method}The
linear fit method is from \citealt{ptvf+92}.} The derived values
and standard errors are $a=1.37\pm 0.05$, $b=2.10\pm 0.21$, the
corresponding $\chi^2=9.97$ for 7 degrees of freedom (dof), and
the possibility $Q$($>\chi^2$)$=0.19$. If we do not include GRB
$011121$, the result gives the same $a$ and $b$, while
$\chi^2=9.67$ for $6$ dof and $Q$($>\chi^2$)$=0.14$. This can be
explained by the large error bar of this GRB, as shown in Figure
\ref{fig:11-breaks}. The mean value of $b$ is consistent with the
theoretical value of $p$ (\citealt{agk+01}). The large scatter of
$b$ is caused by the limitation of the small sample. If we adopt
$a=1.37\pm 0.15$ (3 $\sigma$) and $b=2.2$, the jet break relation
constrains the physical parameters as
\begin{eqnarray}
\epsilon_{e,-1}^{1.2}\epsilon_{B,-3}^{0.8}n^{0.07}E_{j,50.5}^{1.73}\sim
1.1-2.1& &\;(\rm{ISM}),\nonumber\\
\epsilon_{e,-1}^{1.2}\epsilon_{B,-3}^{0.8}A_{\ast,-1}^{0.2}E_{j,50.5}^{1.6}\sim
0.51-1.0& &(\rm{wind}), \label{eqn:universal}
\end{eqnarray}
which implies the \textit{universal} energy reservoir and
relativistic shock physics. The constraints for the ISM case and
the wind case are nearly the same, as long as the typical wind
parameter is relatively small, $A_{\ast}=0.1$
(\citealt{wdhl+03,dw+03,clf+04}). The determination of the
intrinsic mean values of these parameters is needed to combine
equation (\ref{eqn:universal}) with other methods, e.g., the
multiwavelength fits to the overall afterglow light curves.
\cite{pk+01,pk+02} have performed these fits and given the mean
values of $E_{j}\sim 5\times 10^{50}$ ergs, $\epsilon_{e}\sim
0.3$, and $\epsilon_{B}\sim 4\times 10^{-3}$, which are marginally
consistent with the constraints of equation (\ref{eqn:universal}).
We note that \cite{pk+01,pk+02} have assumed the distribution of
initially shocked electrons to be a broken power law. They have
also given a large range of $\epsilon_{B}$, from $10^{-5}$ to
$0.1$. Although the lower and upper limits of $\epsilon_{B}$ are
expected when the relativistic two-stream instability of the
electrons and protons saturates separately (\citealt{ml+99}), the
question is why the same relativistic shock physics leads to very
different $\epsilon_{B}$ values among GRB afterglows. In this
work, we prefer the mean values of physical parameters as
$E_{j}\sim 3\times 10^{50}$ ergs, $\epsilon_{e}\sim 0.1$, and
$\epsilon_{B}\sim 10^{-3}$.

\begin{figure}
\plotone{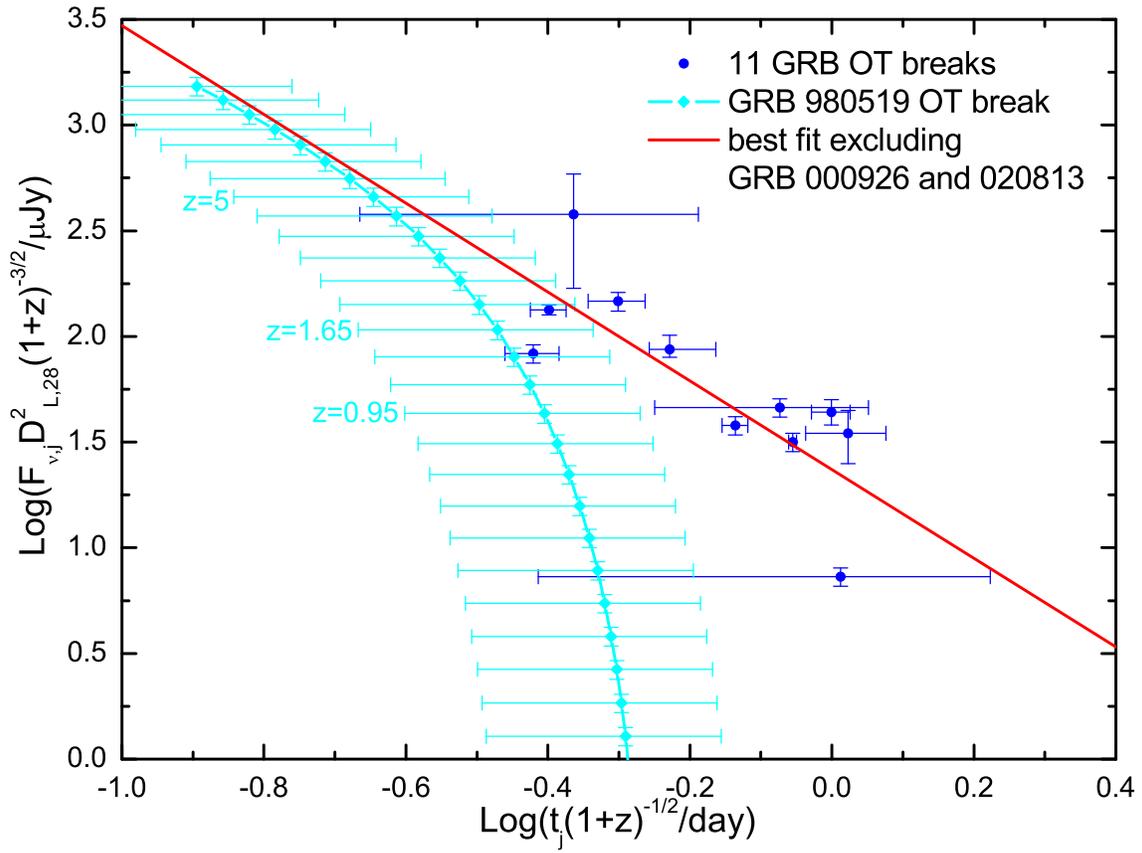} \caption{GRB $980519$ $R$-band jet break data
vary as redshift in the slow-cooling jet break plot. The solid
line with $a=1.37\pm 0.05$, $b=2.10\pm 0.21$ is the best linear
fit, similar to Fig.~\ref{fig:11-breaks}. GRB $980519$ would lie
at $z\gtrsim 1.65$ if it obeys the same jet break relation.
\label{fig:980519break}}
\end{figure}

In Figure \ref{fig:980519break} we calculate the $\mathcal{L_{J}}$
- $\tau_{j}$ plot for GRB $980519$ at different redshift. The line
of GRB $980519$ can be understood as the luminosity distance
$D_{L}$ as a function of $z$, which can be approximated
by\footnotemark\footnotetext{\label{foot:distance}The relative
error is $<2\%$ for $z<2$, $<14\%$ for $z\sim 2-100$, and $<5\%$
for $z>200$. At very small and large $z$, the relative error
approaches to zero. However, we use the exact integral formula of
$D_{L}$ in our calculations.}
\begin{equation}
D_{L}(z)=\frac{c}{H_0}\frac{1+z}{1+0.29z}z,
\end{equation}
for $\Omega_{m}=0.27$ and $\Omega_{\Lambda}=0.73$, where $c$ is
the speed of light. The line begins vertically when $z\ll 1$ and
approaches a diagonal with a slope of $-1$ at $z\gg 1$ in the
$\mathcal{L_{J}}$ - $\tau_{j}$ plot. If GRB $980519$ obeys the
same jet break relation of equation (\ref{eqn:sc-break}), a lower
limit of its redshift, $z\gtrsim 1.65$, is indicated in Figure
\ref{fig:980519break}. This lower limit is consistent with the
nondetection of a supernova signature expected to accompany GRB
$980519$ at late times, which implies $z\gtrsim 1.5$
(\citealt{jhb+01}). However, the redshift determined by the jet
break relation has, in general, two values. The specific jet break
data of GRB $980519$ and the large scatter of $t_{j}$ prevent the
unambiguous determination of its redshift.

\begin{figure}
\plotone{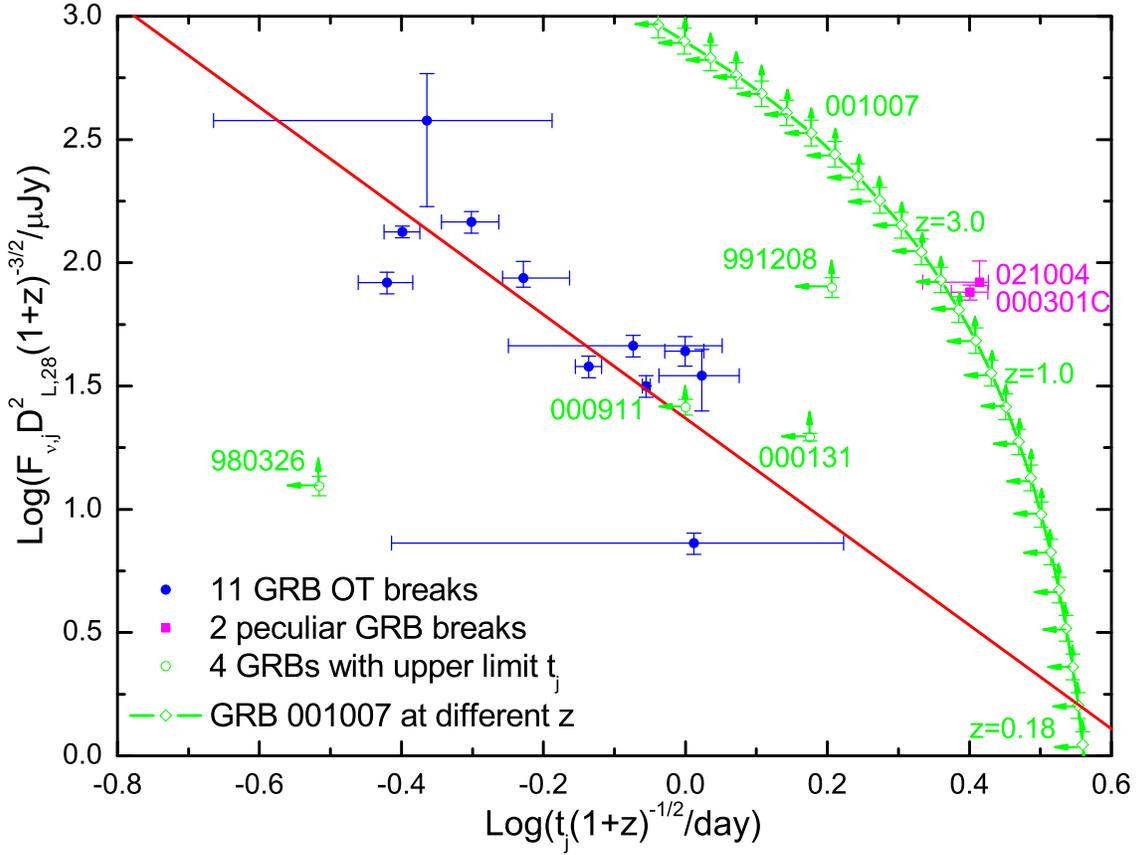} \caption{All $R$-band jet break data of 11
typical and two peculiar GRB OTs and upper limits of $t_{j}$ of
four fast-fading GRB OTs. The solid line with $a=1.37\pm 0.05$ and
$b=2.10\pm 0.21$ is also shown for comparison. Two fast-fading OTs
(GRBs $980326$ and $991208$) and the two peculiar GRB $000301$C
and $021004$ OTs form distinctive classes (less energetic GRB
$980326$ and more energetic bursts for the three others) from the
others. It can be estimated from the jet break relation that the
redshift of the fast-fading GRB $001007$ is $z\sim 0.2$.
\label{fig:all-breaks}}
\end{figure}

Figure \ref{fig:all-breaks} shows the break data of two peculiar
GRBs, GRB $000301$C and $021004$, and five fast-fading GRBs,
together with $11$ typical GRB breaks with known redshifts.
Fluctuations or bumps in the optical light curve before $t_{j}$
will strongly affect the observational determination of $t_{j}$.
However, it is impossible to attribute GRB $000301$C and $021004$
to the universal class obeying the same jet break relation,
because of the uncertainty of their $t_{j}$ and $F_{\nu,j}$. A
rough estimate of the physical parameters assuming the same
$b=2.1$ for these two GRBs will give $a\sim 2.75$, twice the
universal value. Since $a$ is insensitive to the external number
density, and since it is unreasonable to assume the shock physics
will change much in short timescales, we draw an exclusive
conclusion that the jet is re-energized by a factor of $6.3$
($7.3$) for ISM (wind) case, because of the delayed energy
injection from the central engine. \cite{bfk+03} proposed that
fast-fading GRBs belong to the low-energy subclass with respect to
the gamma-ray energy release. However, the case becomes more
complicated in view of the residual energy in the afterglow epoch,
as shown in Figure \ref{fig:all-breaks}. The fast-fading GRB
$000131$ seems to obey the jet break relation, because the line
extrapolating its first detected data to the earlier time $t_{j}$
is nearly parallel to the solid line, since $\alpha\sim b$. GRB
$000911$ also seems to belong to the universal class, if the first
detection time is close to the true $t_{j}$. GRB $980326$ is
identified as a subenergetic GRB afterglow. The spectral indices
of GRBs $991208$ and $001007$ indicate that they belong to the
fast-cooling case, although the indices are not corrected for the
host galaxy extinction, which may make them possibly belong to the
slow-cooling case. For comparison, they are plotted in Figure
\ref{fig:all-breaks}. The redshift of GRB $001007$ is estimated at
$z\sim 0.18$ by assuming that it follows the slow-cooling jet
break relation. However, a reliable redshift can be only estimated
when a large sample of fast-cooling jet break data is achieved.

\section{Conclusions and Discussion}
\label{sec:conclusions} In this paper, we have derived the jet
break relation of GRB optical light curves, $F_{\nu,j}\propto
t_{j}^{-p}$, and given the statistical results of this relation
based on the available sample. Now we summarize our findings and
discuss their implications.

First, the electron distribution index $p=2.10\pm 0.21$ is
achieved in the statistical sense. Conventionally, the late
temporal index $\alpha_2$ or $\alpha$ of fast-fading GRBs is
believed to be the same as $p$. However, there are two caveats on
this assumption. (1) Most importantly, the ambiguity of the
understanding of the sideways expansion of the jet leads to a
great uncertainty of the value of $\alpha_2$. For a non-lateral
spreading jet, $\alpha_2$ is larger than $\alpha_1$ by $3/4$
($1/2$) for ISM (wind) case, because of the deficit of the visual
edge caused by the relativistic beaming effect (\citealt{mr+99}).
GRBs $990123$, $010222$, $020813$, $021004$, and the fast-fading
GRB $000911$ are candidates for nonlateral expansion jets. It
should be pointed out that an explanation with flat electron
spectra of $1<p<2$ fails to account for these bursts, since in
this case $\alpha_2=(p+6)/4$ ($>1.75$) is larger than the observed
value (\citealt{dc+01}). (2) Even though detailed calculation of
sideways expansion evolution results in the light curve of
$\alpha_2\sim p$, there is a tendency for larger $\theta_0$ to
cause relatively flatter $\alpha_2$ (\citealt{hdl+00,wdhm+04}).
There is also an indication that $\alpha_2$ is larger (steeper)
than $p$ in the two-dimensional simulation (\citealt{gmp+01}). In
this case, the jet experiences a lateral expansion stage, while
the emission is mostly arising from the part of the initial
half-opening angle. The temporal index of the steepest light curve
in this case is estimated to be $\alpha_2\sim p+1$. However, the
jet break relation is determined at $t_{j}$, and the only
assumption is $\gamma(t_{j})=\theta_0^{-1}$ (\citealt{fks+01}).
The value of $p=b$ derived from this relation avoids the above
uncertainties and can be used as a better and independent way to
constrain the relativistic shock acceleration physics.

Second, the jet break relation further supports the ``standard
candle'' hypothesis of the afterglows
(\citealt{pk+01,pk+02,fw+01,pkpp+01,bfk+03,bkf+03}). Furthermore,
it also constrains the shock physics to be universal among nine
GRBs in the slow-cooling case (see equation \ref{eqn:universal}).
Since the jet break relation is almost immune from the effect of
external density, it can probe the energy reservoir and shock
physics of GRBs at high redshifts, where the density of the ISM or
the stellar wind may significantly follow the cosmological
evolution (\citealt{ciardi00}). There is another capability of
this relation to distinguish between some peculiar GRBs, e.g., the
GRBs with delayed energy injections before $t_{j}$.

Third, as a by-product, we can estimate the redshift of some jet
break GRBs or fast-fading GRBs by using of the jet break relation,
assuming they follow the standard energy and same shock physics.

The jet break relation itself can not distinguish between the
structures of GRBs jets (\citealt{mrw+98,dg+01,rlr+02,zm+02}).
However, the empirical formula that is used to fit the light
curves gives the required jet break data, e.g., $F_{\nu,j}$,
$t_{j}$, and the break sharpness $s$ (\citealt{bhr+99,rf+01}). The
sharpness $s$ has the potential to probe the jet structure,
because it behaves in a different way as a function of $\theta_0$
(therefore $t_{j}$) in a homogeneous jet and in a structured jet.

The $F_{\nu,j}$ -- $t_{j}$ relation  presents a way to constrain
the value of $p$ and other physical parameters of GRB afterglows.
A more robust result should be based on a larger sample of GRBs
with measured jet breaks in their afterglow light curves, which is
expected in the upcoming \textit{Swift} era.

We thank the referee very much for his/her valuable suggestions
and comments. This work was supported by the National Natural
Science Foundation of China (grants 10233010 and 10221001), the
Ministry of Science and Technology of China (NKBRSF G19990754),
the Natural Science Foundation of Yunnan (2001A0025Q), and the
Research Foundation of Guangxi University.

\clearpage

\begin{deluxetable}{cccccccc}
\tablecolumns{8} \tablewidth{0pc} \tablecaption{$R$-Band Jet Break
Data\label{tab:break}} \tablehead { \colhead {GRB} & \colhead
{$z$} & \colhead {$\alpha_1$} & \colhead {$\alpha_2$} & \colhead
{$t_{j}$} & \colhead {$F_{\nu,j}$} & \colhead {$\beta_{\rm{opt}}$ ($t_{\rm{day}}$)$^{\rm{c}}$} & \colhead {Reference} \\
\colhead {} & \colhead {} & \colhead {} & \colhead {} & \colhead
{(day)}  & \colhead {($\mu$Jy)}  & \colhead {}  & \colhead {} }
\startdata
980519 &\nodata & $1.73\pm 0.04$ & $2.22\pm 0.04$ & $0.55\pm 0.20$  & $31.6\pm 3.16^{\rm{b}}$  & $0.81\pm 0.01(0.45)$ & 1    \\
990123 & 1.6004 & $1.17\pm 0.30$ & $1.57\pm 0.11$ & $1.70\pm 0.22$  & $10.74\pm 3.01$   & $0.75\pm 0.07(<2.8)$ & 2,3  \\
990510 & 1.6187 & $0.46\pm 0.20$ & $1.85\pm 0.26$ & $0.70\pm 0.35$  & $114.54\pm 63.24$ & $0.61\pm 0.12(0.9)$ & 4,3,5  \\
991216 & 1.02   & $     1.0    $ & $     1.8    $ & $1.20\pm 0.40$  & $30\pm 3^{\rm{b}}   $    & $0.58\pm 0.08(1.65)$ & 6,7,8  \\
000301C$^{\rm{a}}$& 2.0404 & $0.72\pm 0.06$ & $2.29\pm 0.17$ & $4.39\pm 0.26$  & $16.23\pm 1.13$   & $\sim 0.6(4.26)$ & 9,10    \\
000926 & 2.0369 & $1.45\pm 0.06$ & $2.57\pm 0.10$ & $1.74\pm 0.11$  & $9.38_{-1.21}^{+1.39}$  & $0.94\pm 0.02(2.26)$ & 11,12  \\
010222 & 1.4768 & $\sim 0.8    $ & $1.57\pm 0.04$ & $0.93_{-0.06}^{+0.15}$  & $30.45_{-2.5}^{+5.1}$  & $0.75(<2.96)$ & 13,14\\
011121 & 0.362  & $1.62\pm 0.39$ & $2.44\pm 0.34$ & $1.20\pm 0.75$  & $22.0\pm 2.2^{\rm{b}}$   & $0.62\pm 0.05(2.5)$ & 15,16\\
011211 & 2.140  & $0.95\pm 0.02$ & $2.11\pm 0.07$ & $1.56\pm 0.02$  & $6.31\pm 0.63^{\rm{b}}$  & $0.61\pm 0.15(<1.52)$ & 17,18,19\\
020405 & 0.6899 & $\sim1.4     $ & $\sim 1.95   $ & $0.95\pm 0.04$  & $50\pm 5^b   $    & $0.74(1.7)$  & 20,21,22\\
020813 & 1.255  & $0.76\pm 0.05$ & $1.46\pm 0.04$ & $0.57\pm 0.05$  & $38.0\pm 3.8^{\rm{b}} $  & $0.93\pm 0.16(0.43)$ & 23,24,25\\
021004$^{\rm{a}}$ & 2.3351 & $0.85\pm 0.01$ & $1.43\pm 0.03$ & $4.74_{-0.80}^{+0.14}$  & $14.675_{-0.445}^{+3.209}$ & $0.39\pm 0.12(5.57)$ & 26,27\\
030326 & 1.986  & $0.77\pm 0.04$ & $1.99\pm 0.06$ & $0.69\pm 0.04$  & $29.65_{-1.59}^{+1.68}$ & $\sim 0.55(0.62)$ & 28,29\\
030329 & 0.1685 & $1.18\pm 0.01$ & $1.81\pm 0.04$ & $0.54\pm 0.05^b$& $3020\pm 302^{\rm{b}}$   & $0.71(0.65)$ & 30,31,32\\
\enddata
\tablecomments{Col.(1) GRB name; col.(2) redshift; col.(3)
temporal decay index ($F_{\nu}\propto t^{-\alpha_1}$) before
$t_{j}$; col.(4) temporal decay index ($F_{\nu}\propto
t^{-\alpha_2}$) after $t_{j}$; col.(5) observed jet break time;
col.(6) flux density at $t_{j}$; col.(7) optical spectrum
$F_{\nu}\propto\nu^{-\beta_{\rm{opt}}}$ at time $t$ in days since
the GRB trigger; and col.(8) references for the redshift and
$R$-band afterglow. $^{\rm{a}}$These two peculiar GRB OTs showed
fluctuations in their early light curves. $^{\rm{b}}$We estimate
$10\%$ uncertainties for these quantities which were not directly
given in the literature. $^{\rm{c}}$The value of
$\beta_{\rm{opt}}$ is sensitive to the assumed host galaxy
extinction correction.} \tablerefs{ (1) \citealt{jhb+01}; (2)
\citealt{kdo+99}; (3) \citealt{hbh+00}; (4) \citealt{vfk+01}; (5)
\citealt{sgk+99}; (6) \citealt{vff+99}; (7) \citealt{hum+00}; (8)
\citealt{gjp+00}; (9) \citealt{jfg+01}; (10) \citealt{rf+01};
(11) \citealt{cdk+00}; (12) \citealt{spm+01}; (13)
\citealt{jpg+01}; (14) \citealt{grb+03}; (15) \citealt{gsw+03};
(16) \citealt{gks+03}; (17) \citealt{fvr+01}; (18)
\citealt{jhf+03}; (19) \citealt{hsg+02}; (20) \citealt{pkb+03};
(21) \citealt{bsf+03}; (22) \citealt{mph+03}; (23)
\citealt{bsc+03}; (24) \citealt{cmt+03}; (25) \citealt{unm+03};
(26) \citealt{mfh+02}; (27) \citealt{hwf+03}; (28)
\citealt{ggk+03}; (29) \citealt{psa+04}; (30) \citealt{gpe+03};
(31) \citealt{sks+03}; (32) \citealt{mgs+03};}
\end{deluxetable}

\begin{deluxetable}{ccccccc}
\tablecolumns{7} \tablewidth{0pc} \tablecaption{$R$-Band Data of
Fast-fading GRB Afterglows\label{tab:fastfading}} \tablehead {
\colhead {GRB} & \colhead {$z$} & \colhead {$\alpha$} & \colhead
{$t_{j}$} & \colhead {$F_{\nu,j}$} & \colhead {$\beta_{\rm{opt}}$ ($t_{\rm{day}}$)} & \colhead {Reference} \\
\colhead {} & \colhead {} & \colhead {} & \colhead {(days)}     &
\colhead {($\mu$Jy)}  & \colhead {}   & \colhead {} } \startdata
980326 & $\sim0.9$& $2.0\pm 0.1$   & $<0.42$  & $>10.09_{-0.89}^{+0.97}$ & $0.8\pm 0.4(2.38)$ & 1\\
991208 & $0.7063$ & $2.30\pm 0.07$ & $<2.1$   & $>100_{-8.8}^{+9.6}$     & $1.05\pm 0.09(3.8)^{\rm{a}}$ & 2,3,2\\
000131 & $4.500$  & $2.25\pm 0.19$ & $<3.513$ & $>1.5_{-0.054}^{+0.056}$ & $\sim 0.70(3.5)$ & 4\\
000911 & $1.0585$ & $1.46\pm 0.05$ & $<1.435$ & $>15.85_{-1.14}^{+1.23}$ & $0.75\pm 0.01(>1.44)$ & 5,6\\
001007 & \nodata  & $2.03\pm 0.11$ & $<3.95$  & $>19.05_{-2.15}^{+2.42}$ & $1.24\pm 0.57(3.94)^{\rm{a}}$ & 7\\
\enddata
\tablecomments{Col.(1) GRB name; col.(2) redshift; col.(3)
temporal decay index ($F_{\nu}\propto t^{-\alpha}$); col.(4)
observed jet break time; col.(5) flux density at $t_{j}$; col.(6)
optical spectrum $F_{\nu}\propto\nu^{-\beta_{\rm{opt}}}$ at time
$t$ in days since the GRB trigger; and col.(7) references for the
redshift and $R$-band afterglow. $^{\rm{a}}$The value is not
de-reddening for the extinction of the host galaxy and the actual
one may be flatter.} \tablerefs{ (1) \citealt{bkd+99}; (2)
\citealt{csg+01}; (3) \citealt{jhp+99}; (4) \citealt{ahp+00}; (5)
\citealt{pbk+02}; (6) \citealt{lcg+01}; (7) \citealt{ccg+02}; }
\end{deluxetable}

\clearpage

\end{document}